\documentclass[aps,prd,
preprintnumbers,groupedaddress]{revtex4}
\usepackage{epsbox}
\usepackage{amsmath}
\usepackage[dvips]{graphicx}

\begin{document}
\preprint{\vtop{
{\hbox{YITP-07-36}\vskip-0pt
                 \hbox{KANAZAWA-07-07} \vskip-0pt
}
}
}


\title{ 
 A new tetra-quark interpretation of ${\bf X(3872)}$
}

\author{
Kunihiko Terasaki   
}
\affiliation{
Yukawa Institute for Theoretical Physics, Kyoto University,
Kyoto 606-8502, Japan \\
Institute for Theoretical Physics, Kanazawa University, 
Kanazawa 920-1192, Japan
}

\begin{abstract}
{
A new tetra-quark interpretation of $X(3872)$ is presented. In this 
model, $X(3872)$ consists of two degenerate tetra-quark mesons 
$\sim\{[cn](\bar c\bar n)\pm (cn)[\bar c\bar n]\}_{I=0}$, and, 
therefore, it is naturally understood that $X(3872)$ decays into two 
different eigenstates of $G$ parity. 
}
\end{abstract}

\maketitle

A narrow charmonium-like resonance, which is called $X(3872)$, has been 
observed in the $\pi^+\pi^-J/\psi$ mass distribution~\cite{Belle-X-rho} 
from the $B^+\rightarrow K^+\pi^+\pi^-J/\psi$ decay, and its existence 
has been confirmed by the CDF~\cite{CDF-X}, D0~\cite{D0-X} and 
Babar~\cite{Babar-X} collaborations. Its mass and width have been 
compiled as 
$m=3871.2 \pm 0.5$ MeV and $\Gamma < 2.3$ MeV, with 
$CL=90$ \%~\cite{PDG06}. 
Because the observed dipion mass spectrum is concentrated at high 
values, it was speculated~\cite{Belle-X-rho} that the decay might 
proceed through the $X(3872)\rightarrow \rho^0J/\psi$ reaction. However, 
this has not yet been established conclusively, because a 
search~\cite{Babar-charged-partner} for its charged partners in 
$B\rightarrow K\pi^-\pi^0J/\psi$ decays  has given a negative result 
and because the $X(3872)\rightarrow \pi^0\pi^0J/\psi$ decay has not yet 
been searched for. In addition, another resonance peak has been observed 
at the same mass in the $\pi^+\pi^-\pi^0J/\psi$ channel. By identifying 
these two resonances, the ratio of the measured rates has been 
determined as~\cite{Belle-X-omega} 
$ 
Br(X(3872)\rightarrow \pi^+\pi^-\pi^0J/\psi)/
Br(X(3872)\rightarrow \pi^+\pi^-J/\psi)
=1.0\pm 0.4\pm 0.3
$.  
This implies that conservation of $G$-parity and 
the above identification are incompatible. From the $\pi^+\pi^-\pi^0$ 
mass distribution in $X(3872)\rightarrow \pi^+\pi^-\pi^0J/\psi$, it 
has also been   
speculated that it proceeds through the sub-threshold decay 
$X(3872)\rightarrow \omega J/\psi$~\cite{Belle-X-omega}. 
However, if the parents of these decays were the same meson $X(3872)$, 
as implicitly assumed, such a process would imply that isospin 
conservation in strong interactions of $X(3872)$ is badly 
broken~\cite{Suzuki},    
$| 
{A(X(3872)\rightarrow \rho^0J/\psi)}/
                        {A(X(3872)\rightarrow \omega J/\psi)}|   
=0.27\pm 0.02$,        
in contrast with known strong interactions. The same 
experiment~\cite{Belle-X-omega} observed the 
$X(3872)\rightarrow \gamma J/\psi$ decay. From this result, it has been 
speculated that the charge conjugation parity of $X(3872)$ is even, 
where, again, it has been implicitly assumed that $X(3872)$ is a single 
meson state. Regarding its spin and parity, the angular analysis favors 
$J^P=1^+$ over other quantum numbers~\cite{Belle-X-J^P}. 
In the $B\rightarrow D^0\bar D^0\pi^0K$ decay, an extra resonance peak 
has been observed at the $D^0\bar D^0\pi^0$ 
threshold~\cite{Belle-D-Dbar-pi}. It can be interpreted as an $X(3872)$ 
signal~\cite{Hanhart}, although its measured mass,  
$\displaystyle{M=(3875.9\pm 0.7^{+0.3}_{-1.6}\pm 0.4)}$ MeV,  
is slightly larger than that of $X(3872)$ mentioned above. The measured 
rate for the decay $X(3872)\rightarrow D^0\bar D^0\pi^0$ is 
larger by about an order of magnitude than that for the 
$X(3872)\rightarrow \pi^+\pi^-J/\psi$ decay. 

To understand $X(3872)$, various theoretical models, for example, 
a loosely bound molecular state~\cite{molecule}, a diquark-antidiquark 
$[cn][\bar c\bar n]$ (where $n=u,\,d$) state~\cite{Maiani}, a hybrid 
meson~\cite{hybrid}, a glueball with a $c\bar c$ 
admixture~\cite{glue-ball}, etc., 
in addition to the conventional charmonium~\cite{BG}, have been 
proposed. However, it seems difficult to reconcile $X(3872)$ with 
the above charmonium with regard to mass~\cite{Zhu}. 
Among the abovementioned exotic models, the $D^0\bar D^{*0}$ 
molecule~\cite{molecule} might easily explain the violation of isospin 
conservation speculated above, because the $D^+D^{*-}$ molecule, as 
the counterpart of $D^0\bar D^{*0}$, is not included. However, the 
absence of such a state leads to a problem in the production of 
$X(3872)$; specifically, the molecular model predicts 
$\mathcal{R}_{\rm molecule}< 0.1$~\cite{BK}, while experiments have 
found  
$ 0.13 < \mathcal{R}_{\rm exp} < 1.10 $ at 90 \% CL~\cite{Babar-R},   
where 
$\mathcal{R} \equiv 
{Br(B^0\rightarrow X(3872)K^0)}/ 
            {Br(B^+\rightarrow X(3872)K^+)}$. 
In contrast, the diquark-antidiquark model~\cite{Maiani}, in which 
$X(3872)$ is assigned to a $[cn][\bar c\bar n]$ (where $n=u,\,d$) 
predicts that $\mathcal{R}_{[cn][\bar c\bar n]} = 1$. In this sense, 
$\mathcal{R}$ seems to favor the diquark-antidiquark model rather than 
the $D^0\bar D^{*0}$ molecule. However, the diquark-antidiquark 
model~\cite{Maiani} has predicted a large difference between the masses 
of $X_d\sim [cd][\bar c\bar d]$ and $X_u\sim [cu][\bar c\bar u]$, which 
are produced in the decays of $B^0$ and $B^+$, respectively, i.e., 
$m_{X_d} - m_{X_u} \simeq (7 \pm 2)$ MeV.  
This result is larger than the measured $(2.7\pm 1.3\pm 0.2)$ 
MeV~\cite{Babar-R}. These experimental results suggest that isospin 
symmetry is compatible with the production of $X(3872)$. In addition, 
the diquark-antiquark model predicts the existence of its charged 
partners, in contradiction with the negative result obtained from the 
experimental search, as mentioned above. In the models listed above, 
$X(3872)$ is assigned to a single meson state with a definite 
$G$-parity and thus they encounter the serious problem that the strong 
interactions of $X(3872)$ do not conserve $G$-parity, in contrast with 
the known ones. Thus, all the existing models seem to have some serious 
problem. Experimental data on $X(3872)$ and its theoretical 
interpretations are reviewed, for example, in Refs.~\cite{Zhu} and 
\cite{review}. 

Before introducing a new four-quark interpretation of $X(3872)$, we 
very briefly review four-quark mesons. They can be classified into four 
groups~\cite{Jaffe},
\begin{equation}
\{qq\bar q\bar q\} 
= [qq][\bar q\bar q] \oplus (qq)(\bar q\bar q) 
        \oplus \{[qq](\bar q\bar q) \pm (qq)[\bar q\bar q]\},
                                                 \label{eq:4-quark}
\end{equation}
where the parentheses and the square brackets denote symmetry and 
anti-symmetry, respectively, of the wavefunction under the exchange of 
flavors between them. Each term on the right-hand side (r.h.s.) of 
Eq.~(\ref{eq:4-quark}) is again classified into two classes, because 
there are two different ways to construct a color singlet 
$\{qq\bar q\bar q\}$ state, i.e., to take 
${\bf \bar 3_c}\times{\bf 3_c}$ and ${\bf 6_c}\times {\bf \bar 6_c}$ 
of the color $SU_c(3)$. Although these two can mix with each other in 
general, here we ignore such  mixing for simplicity. The allowed spins 
of low-lying four-quark mesons in the flavor symmetry limit are listed 
in Table~I. As seen in the table, the $[qq][\bar q\bar q]$ mesons with 
${\bf \bar 3_c}\times{\bf 3_c}$ have $J^P=0^+$. When $q=u,\,d,\,s$, 
they accurately describe the observed low-lying scalar 
mesons~\cite{PDG06}, $a_0(980)$, $f_0(980)$, $\kappa$ and $f_0(600)$, 
as suggested in Ref.~\cite{Jaffe} and supported recently by lattice 
QCD results~\cite{Liu}. However, the corresponding scalar, axial-vector 
and tensor mesons which arise from the ${\bf 6_c}\times{\bf \bar{6}_c}$ 
component, as seen in Table~I, have not yet been observed near the 
scalar mesons. This implies 
that the ${\bf 6_c}\times{\bf \bar{6}_c}$ state cannot be bound, or it 
would be much heavier (even if it could be bound) than the 
${\bf \bar 3_c}\times{\bf 3_c}$, because the forces between $qq$ (and 
$\bar q\bar q$) are repulsive when $qq$ (and $\bar q\bar q$) are of 
$\bf{6_c}$ (and $\bf{\bar{6}_c}$)~\cite{Hori}. As a result, the mixing 
between the ${\bf\bar 3_c}\times{\bf 3_c}$ and 
${\bf 6_c}\times{\bf\bar{6}_c}$ states might be small, and therefore, 
it is conjectured that ignoring the above mixing introduces only a 
small error. 

When one of the light quarks in $[qq][\bar q\bar q]$ is replaced by the 
charm quark, $c$, open-charm scalar mesons can be obtained, and the 
well-known $D_{s0}^+(2317)$~\cite{PDG06} has been successfully assigned 
to the iso-triplet 
$\hat F_I^+\sim [cn][\bar s\bar n]_{I=1}$~\cite{Terasaki-D_{s0}}.   
(Our notation for open-charm scalar mesons is defined in 
Ref.~\cite{Terasaki-D_{s0}}.) In fact, by adopting this assignment, its 
narrow width can be easily understood, because of the small overlap of 
the color and spin wavefunctions~\cite{Terasaki-ws}, and the 
experimental constraint~\cite{CLEO-D_{s0}},   
\begin{equation}
\frac{\Gamma(D_{s0}^+(2317) \rightarrow D_{s}^{*+}\gamma)}  
       {\Gamma(D_{s0}^+(2317) \rightarrow D_{s}^{+}\pi^0)} 
< 0.059, 
\end{equation}
can be naturally satisfied~\cite{HT,Trento} in the approach in which 
the measured ratio of decay rates~\cite{PDG06},  
\begin{equation}
\frac{\Gamma(D_{s}^{*+} \rightarrow D_{s}^{+}\pi^0)}
  {\Gamma(D_{s}^{*+} \rightarrow D_{s}^{+}\gamma)}
=0.062\pm 0.008,  
\end{equation}
is reproduced. 
The above ratio implies that isospin non-conserving interactions in the 
charm-strange system are much weaker than the electromagnetic ones, 
which are much weaker than the isospin conserving strong interactions; 
i.e., isospin conservation is approximately realized in the open-charm 
system, as in known strong interactions. In addition, it has been 
discussed~\cite{Terasaki-production} that the iso-singlet $\hat F_0^+$ 
might have already been observed in the radiative channel   
$B\rightarrow \bar DD_s^{*+}\gamma$~\cite{Belle-D_{s0}}, and it has 
been conjectured that the neutral $\hat F_I^{0}$ and doubly charged 
$\hat F_I^{++}$ partners of $D_{s0}^+(2317)$ will be found in hadronic 
weak decays of $B$ mesons. 
\begin{center}
\begin{table}[t]
\caption{ 
Spins of four-quark $\{qq\bar q\bar q\}$ mesons in the flavor symmetry 
limit. 
}
\vspace{1mm}
\begin{center}
\begin{tabular}{c | c | c | c | c | c | c}
\hline
Flavor &\multicolumn{2}{|c|}{$[qq][\bar q\bar q]$}
&\multicolumn{2}{|c|}{$(qq)(\bar q\bar q)$}  
&\multicolumn{2}{|c}{$[qq](\bar q\bar q) \pm (qq)[\bar q\bar q]$}
\\
\hline
Color &${\bf \bar 3_c}\times{\bf 3_c}$
&${\bf 6_c}\times{\bf \bar 6_c}$
&${\bf \bar 3_c}\times{\bf 3_c}$
&${\bf 6_c}\times{\bf \bar 6_c}$
&${\bf \bar 3_c}\times{\bf 3_c}$
&${\bf 6_c}\times{\bf \bar 6_c}$
\\
\hline
Spin
& 0 & $0\oplus 1\oplus 2$ & $0\oplus 1\oplus 2$ & 0 & 1 & 1 
\\
\hline
\end{tabular}\vspace{-3mm}
\end{center}
\end{table}  
\end{center}
\vspace{-0mm}

Although $D_{s0}^+(2317)$ is a natural and feasible possibility for the 
iso-triplet scalar four-quark meson $\hat F_I^+$ with 
$\bf{\bar 3_c}\times\bf{3_c}$, as seen above, its straightforward 
extension to hidden-charm axial-vector mesons involves some problems, 
as discussed above. The $(cn)(\bar c\bar n)$ system corresponding to 
the second term on the r.h.s. of Eq.~(\ref{eq:4-quark}) can have 
$J^{PC}=1^{++}$ for ${\bf \bar 3_c}\times{\bf 3_c}$. However, it is not 
clear whether it can be bound. For example, if $(nn)(\bar n\bar s)$ 
were bound, there would exist strange scalar mesons with $I=3/2$, and 
hence the $\pi K$ phase shift with $I=3/2$ should pass through $\pi/2$ 
at the resonant energy. However, no indication of such phenomena has 
been observed~\cite{pi-K}. In addition, if $X(3872)$ were assigned 
to $(cn)(\bar c\bar n)_{I=0}$, its $G$-parity would be even, as long as 
isospin is a good quantum number. Therefore, it is difficult to avoide 
the problem of the non-conservation of $G$-parity, because $X(3872)$ 
decays into the $\pi\pi J/\psi$ and $\pi\pi\pi J/\psi$ states with 
opposite $G$-parities. 

Now we propose a new interpretation of $X(3872)$. The last two 
components of Eq.~(\ref{eq:4-quark}), and hence the corresponding 
$[cq](\bar c\bar q)$ and $(cq)[\bar c\bar q]$, which belong to the 
ideally mixed ${\bf \overline{60}_f}$ and ${\bf{60}_f}$ multiplets, 
respectively, of the flavor $SU_f(4)$, have $J^{P}=1^{+}$, as seen in 
Table~I. Although the light-quark sector of 
$\{[qq](\bar q\bar q) \pm (qq)[\bar q\bar q]\}$ axial-vector mesons 
have been studied since long ago~\cite{Jaffe}, they have not been 
identified with any observed axial-vector mesons since it is difficult 
to definitively assign observed axial-vector mesons to the members of 
this class of mesons. This is because their quantum numbers (flavors, 
isospin, $G$-parity, etc.) are included in the conventional $q\bar q$ 
and $(qq)(\bar q\bar q)$ axial-vector mesons (and $[qq][\bar q\bar q]$ 
states with ${\bf 6_c}\times{\bf \bar{6}_c}$ if they can be bound), 
and because these mesons with the same quantum numbers might mix with 
each other through the common final states of their decays. 

We now consider the hidden-charm sector. Because 
$X(\overline{60}_f)\sim [cn](\bar c\bar n)_{I=0}$ 
and $X(60_f)\sim (cn)[\bar c\bar n]_{I=0}$ 
can be connected with each other under charge conjugation, they can 
have equal masses as long as charge-conjugation parity is conserved. 
However, neither of them can be an eigenstate of $G$-parity, and hence 
they mix with each other to form eigenstates of $G$-parity, i.e., 
$X(\pm)\sim X({\overline{60}_f})\pm X({60_f})$, under $G$-parity 
conserving strong interactions. In this case, $c\bar cu\bar u$ and 
$c\bar cd\bar d$ are included with equal weight in $X(\pm)$, so that 
the ratio $\mathcal{R}$ discussed above is expected to be 
$\mathcal{R}=1$, which is consistent with experiment. The point here is 
whether the mass difference $|\Delta m_X|=|m_{X(-)} - m_{X(+)}|$ caused 
by the above mixing is sizable. If it is sizable, $X(\pm)$ will appear 
as two different states with different masses, and thus they could not 
be identified with $X(3872)$. However, if it is not sizable, then 
$X(\pm)$ will appear as a single meson state, $X(3872)$, with regard to 
mass, while they will behave as two different states in their decays, 
because of their different $G$-parities. 

The mass difference $|\Delta m_X|$ caused by the mixing of hidden-charm 
tetra-quark mesons through common final states of their decays is 
expected to be small, because interactions causing their decays are 
weak, due to the small overlap of the color and spin wavefunctions, as 
discussed below, in contrast with the case of the light quark sector 
with smaller mass. This is similar to the result obtained in 
Refs.~\cite{Trento} and ~\cite{HT} that the rate for the 
$D_{s0}^+(2317)\rightarrow D_s^+\pi^0$ decay at a higher energy scale 
is small, while the rate for $a_0(980)\rightarrow \eta\pi$ at a lower 
energy scale is large. With regard to this, it should be noted that a 
chiral model in the framework of broken $SU_f(4)$ symmetry~\cite{Oset} 
predicts two axial-vector mesons with opposite $G$-parities near 
$X(3872)$. Their mass difference is small (3 MeV), and it seems to be 
within energy resolutions of existing experiments on $X(3872)$. These 
axial-vector meson states could be realized by $X(\pm)$, as seen above. 
Therefore, by assuming that $X(3872)$ consists of $X(-)$ and $X(+)$, 
it could be understood that $X(3872)$ acts like a single meson state 
with regard to mass, while the former decays into $\pi\pi J/\psi$ and 
the latter into $\pi\pi\pi J/\psi$ without violating the usual 
$G$-parity conservation.  

The narrow width of $X(3872)$ can be understood as resulting from a 
small overlap of the color and spin wavefunctions, as in the case of 
$D_{s0}^+(2317)$~\cite{Terasaki-ws,Trento,HT}. Its kinematically 
allowed two-body decays are $X(3872)\rightarrow \eta J/\psi$ and 
$\eta_c\omega$. Although they have not yet been observed, they could be 
dominant decays of $X(3872)$, because they can be $S$-wave decays. In 
fact, the measured branching fraction for the decay 
$X(3872)\rightarrow D^0\bar D^0\pi^0$, 
which can proceed through the $S$-wave decay 
$X(3872)\rightarrow D^0"\bar D^{*0}" + "D^{*0}"\bar D^{0}$,  
is larger by about an order of magnitude than the measured   
$Br(X(3872)\rightarrow \pi^+\pi^-J/\psi)$, as mentioned above. Here 
$"D^{*0}"$ represents the virtual or kinematically allowed part of the 
$D^{*0}$ resonance. If the decay $X(3872)\rightarrow \pi^+\pi^-J/\psi$ 
proceeds through $X(3872)\rightarrow \rho^0J/\psi$, as speculated by 
the Belle collaboration, it must be strongly suppressed, due to the 
isospin non-conservation. If it proceeds through 
$X(3872)\rightarrow f_0(600)J/\psi$, it must be a $P$-wave decay, so 
that it would be suppressed relatively to its $S$-wave decays. 

With this in mind, we consider 
$X(3872)\rightarrow D^0"\bar D^{*0}" + "D^{*0}"\bar D^{0}$ 
as an example, and decompose 
$X_u(\overline{60}_f)\sim [cu](\bar c\bar u)$ into a sum of products of 
$\{c\bar u\}$ and $\{u\bar c\}$ pairs. Then, we have 
\begin{equation}
\hspace{-10mm}
|{[cu]_{\bar 3_c}^{1_s}(\bar c\bar u)_{3_c}^{3_s}}\rangle_{1_c}^{3_s}
= \frac
{1}{2}\sqrt{\frac{1}{3}}
|{\{c\bar u\}_{1_c}^{1_s}    
\{u\bar c\}_{1_c}^{3_s}}\rangle_{1_c}^{3_s} 
+ 
\frac
{1}{2}\sqrt{\frac{1}
{3}}
|{\{c\bar u\}_{1_c}^{3_s}\{u\bar c\}_{1_c}^{1_s}}\rangle_{1_c}^{3_s} 
+ \cdots.
                                            \label{eq:decomposition}
\end{equation}
The coefficient of the first term on the r.h.s. in 
Eq.~(\ref{eq:decomposition}) provides the overlap of the color and spin 
wavefunctions between $X_u(\overline{60}_f)$ and $D^0"\bar D^{*0}"$. 
It is small, as in the case of $D_{s0}^+(2317)\rightarrow D_s^+\pi^0$. 
In the same way, a small overlap of the color and spin wavefunctions in 
the $\eta J/\psi$ and $\eta_c\omega$ decays of $X(3872)$ can be obtained. 
Here we recall that the rate for the 
$D_{s0}^+(2317)\rightarrow D_s^+\pi^0$ decay is small, because of the 
small overlap of the color and spin wavefunctions, although the phase 
space volume is not necessarily small~\cite{Terasaki-ws,Trento,HT}. 
Similarly, the rates for the $X(3872)\rightarrow \eta J/\psi$ and 
$\eta_c\omega$ decays are expected to be small. The rate for the 
$X(3872)\rightarrow D^0"\bar D^{*0}" + "D^{*0}"\bar D^{0}$ decay should  
be smaller than the above decays, since its phase space volume is much 
smaller. In this way, we can understand the narrow width of $X(3872)$, 
although we cannot predict its value at the present stage, because we 
have no useful input data. 

In addition, the present scheme is attractive, because it contains a 
rich spectrum of axial vector mesons. Among open-charm mesons, 
charm-strange axial-vector mesons, which are mixtures of 
$D_{s1}(\overline{60}_f)_{I,0}\sim [cn](\bar{s}\bar{n})_{I=1,0}$ 
and $D_{s1}({60}_f)_{I,0}\sim (cn)[\bar{s}\bar{n}]_{I=1,0}$,  
can exist, according to this model. There is reason to believe that the 
observed $D_{s1}^+(2460)$~\cite{CLEO-D_{s0}} might be one of these $I=1$ 
members, because the measured small ratio of decay rates~\cite{PDG06}, 
$\Gamma(D_{s1}^+(2460)\rightarrow D_s^{+}\gamma)/
\Gamma(D_{s1}^+(2460)\rightarrow D_s^{*+}\pi^0) 
= 0.31 \pm 0.06$, 
can be easily understood in this assignment, as in the case of 
$D_{s0}^+(2317)$~\cite{HT,Trento}. By contrast, if $D_{s1}^+(2460)$ were 
assigned to an iso-singlet state (the conventional $\{c\bar s\}$ or 
an iso-singlet four-quark meson), it would be difficult to naturally 
understand the above ratio, because the isospin non-conserving 
interactions are much weaker than the electromagnetic ones, as seen 
above. In addition to the above charm-strange mesons, we 
can have various exotic axial-vector mesons with $C=\pm 2$, 
$C=-S=\pm 1$, hidden-charm strange mesons, etc., as well as 
$X_I(\mp)\sim \{[cn](\bar c\bar n)\pm (cn)[\bar c\bar n]\}_{I=1}$ 
and $X^s(\pm)\sim \{[cs](\bar c\bar s)\pm (cs)[\bar c\bar s]\}$ 
as the partners of $X(\pm)$ in the present scenario, where $C$ and $S$ 
are the charm and strangeness quantum numbers, respectively. They will 
be studied elsewhere. 

In summary, we have studied the newly discovered resonance $X(3872)$ 
and discussed the fact that all the existing models of $X(3872)$ have 
some serious problems. Then, we presented the new tetra-quark 
interpretation that $X(3872)$ consists of two iso-singlet tetra-quark 
mesons, 
$X(\pm)\sim\{[cn](\bar c\bar n)\pm (cn)[\bar c\bar n]\}_{I=0}$, 
with opposite $G$ parities. In this scenario, we do not need to assume 
a large violation of isospin and $G$-parity conservation in the 
strong interactions of $X(3872)$. 

\section*{Acknowledgments} 
The author would like to thank the members of the Yukawa Institute for 
Theoretical Physics at Kyoto University. This work was motivated by 
discussions during the workshop YKIS2006 on "New Frontiers on QCD". He 
also would like to thank Professor E.~Oset and Mr. D.~Gamermann, IFIC, 
Centro Mixto Universidad de Valencia-CSIC, for valuable discussions and 
the nuclear theory group of the RCNP, Osaka University, for their 
hospitality during his stay. 


\end{document}